\documentclass[twocolumn]{aastex631}
\usepackage{newtxtext,newtxmath}

\usepackage{graphicx}
\submitjournal{AJ}

\shorttitle{Analysis of eclipsing binary KIC 7023917}
\shortauthors{Gajdo\v{s} et al.}

\begin{document}

\title{Analysis of KIC 7023917 -- spotted low-mass ratio eclipsing binary with $\delta$ Scuti pulsations}

\author[0000-0003-1478-3256]{Pavol Gajdo\v{s}}
\affiliation{Astronomical Institute, Czech Academy of Sciences, Fri\v{c}ova 298, 25165, Ond\v{r}ejov, Czech Republic}
\affiliation{Institute of Physics, Faculty of Science, Pavol Jozef \v{S}af\'arik University, 04001 Ko\v{s}ice, Slovakia}
\email{pavol.gajdos@asu.cas.cz, pavol.gajdos@upjs.sk}

\author[0000-0002-1516-1216]{\v{S}tefan Parimucha}
\affiliation{Institute of Physics, Faculty of Science, Pavol Jozef \v{S}af\'arik University, 04001 Ko\v{s}ice, Slovakia}

\author[0000-0002-7602-0046]{Marek Skarka}
\affiliation{Astronomical Institute, Czech Academy of Sciences, Fri\v{c}ova 298, 25165, Ond\v{r}ejov, Czech Republic}

\author{Mat\'{u}\v{s} Kamenec}
\affiliation{Institute of Physics, Faculty of Science, Pavol Jozef \v{S}af\'arik University, 04001 Ko\v{s}ice, Slovakia}

\author{Jozef Lipt\'{a}k}
\affiliation{Astronomical Institute, Czech Academy of Sciences, Fri\v{c}ova 298, 25165, Ond\v{r}ejov, Czech Republic}
\affiliation{Astronomical Institute, Faculty of Mathematics and Physics, Charles University, V Hole\v{s}ovi\v{c}k\'{a}ch 2, 180 00 Praha 8, Czech Republic}

\author[0000-0002-2656-909X]{Raine Karjalainen}
\affiliation{Astronomical Institute, Czech Academy of Sciences, Fri\v{c}ova 298, 25165, Ond\v{r}ejov, Czech Republic}

\begin{abstract}
Times of minima of eclipsing binary KIC\,7023917 show quasiperiodic anti-symmetric deviations from the calculated one with an amplitude of up to 10 minutes and a period of 200~--~300 days. These changes correlate with the observed variations of the light-curve maxima (amplitude and phase separation). 
We used photometric data obtained by \textit{Kepler} and \textit{TESS} missions to analyse the times of minima and determine system parameters. The phases and amplitudes of the maxima were measured to study the O'Connell effect. As an additional source of information, we performed ground-based multi-colour photometric observation and determined the radial velocities of the system from our spectroscopic measurements.
We could explain long-term variations of the light-curve shape and times of the eclipses using the cold star spot located on the secondary component and the modification of its size. Based on our modelling, the system consists of a primary main-sequence star of spectral type A7 and an evolved, oversized secondary component with a mass ratio of only 0.1 due to past mass transfer. Calculation of absolute parameters gives us the mass of the primary component about 1.8~M$_\sun$ and 0.2~M$_\sun$ for the secondary one, and radii of 2.2~R$_\sun$ of the primary star and 0.9~R$_\sun$ of secondary one, respectively. The studied low-mass ratio eclipsing binary is probably a progenitor of the variable star of EL~CVn type. A multiple-period photometric variability was disclosed in the \textit{TESS} data ranging from half to two hours due to $\delta$~Scuti-type pulsations of the primary component.
\end{abstract}

%Stars: binaries: eclipsing -- Stars: starspots -- Stars: oscillations -- Techniques: photometric -- Techniques: radial velocities -- Stars: individual:  KIC\,7023917}

\section{Introduction}

Eclipsing binary stars (EB) are an important group of variable stars that give us the unique possibility to determine the physical parameters of the components by studying their light curves (LC) and mutual changes in radial velocities (RV). Knowledge of this set of parameters (such as mass, radius, and effective temperature) is crucial for theoretical models of star evolution, statistical studies, and stellar astrophysics. In addition, some other interesting features (pulsations, star spots, or changes in times of eclipses) are commonly observed in the LCs of EBs.

The International Variable Star Index (VSX) catalogue of variable stars now includes nearly one million EBs \citep{Watson2006}. Almost 3\,000 EBs are located in the \textit{Kepler} field and have been properly studied using data from this space mission. They are listed in the Kepler Eclipse Binary Catalog \citep{Kirk2016}. Almost 2.5 million new EBs were detected by the \textit{Gaia} mission \citep{Mowlavi2023}. High-precision photometry from space missions \textit{Kepler}, \textit{Kepler-K2}, and \textit{TESS} allows us to study the finest effects in these systems on long uninterrupted time scales.

One such interesting system is the \object[KIC 7023917]{KIC\,7023917} (GSC\,03129-01771, TYC\,3129-1771-1, TIC\,158794976). It was mentioned as a possible eclipsing binary in a catalogue of \textit{ASAS3-North} project \citep[with denotation as ASAS\,J191452+4230.1;][]{Pigulski2008}. The mean brightness is \textit{V}$\sim$10.10. The spectral type of the primary component was estimated based on the \textit{LAMOST} spectra as A7V \citep{Frasca2016}. In Gaia Data Release 3 (DR3), the basic parameters are as follows: temperature $7461\pm15$~K, metallicity $-0.4021\pm0.0130$~dex, and surface gravity $3.955\pm0.0046$~dex \citep{Gaia2023}. \textit{Gaia} mission also measured parallax of this star to $2.337\pm0.012$~mas, corresponding to distance of $427.8\pm2.2$~pc.

We focused on analysing the LC of the EB KIC\,7023917 collected by space missions \textit{Kepler} and \textit{TESS} and studying observed additional effects on it. Because the system shows significant variations in minima times, our main motivation was to describe them. We collected ground-based photometric and spectroscopic data which gives us supplementary information to the space-based ones. 

From the performed analysis, the system belongs to a rare group of eclipsing binaries with a low mass ratio, making it an intriguing object of study. Notably, the secondary component is significantly oversized for its mass, which is an unusual characteristic that points to a complex evolutionary history. The oversized secondary indicates that the system has undergone an extensive mass-transfer phase. Currently, the secondary star starts to contract and increase its effective temperature without substantial changes in luminosity. This contraction phase is a transitional period before the secondary component will eventually finish as a white dwarf in EL~CVn-type binary \citep{Wang2020}. Similar systems with extremely low mass ratios and such dramatic evolutionary paths challenge the theoretical models of stellar and binary star evolution -- mass transfer processes, stellar structure, and the end stages of stellar evolution. The study of these systems is crucial for refining models and improving our knowledge of the complex interactions and life cycles of binary stars.
%Period analysis reveals $\delta$~Scuti pulsations with a period of 50 minutes.

This paper is organized as follows. In Section~\ref{lc}, we present the light curve of this binary with its basic parameters and changes in its shape. Section~\ref{spec} is dedicated to our spectroscopic observations and determinations of radial velocities. Multi-colour ground-based photometry is described in Sec.~\ref{phot}. The analytical model of EB is given in Sec.~\ref{model}. We analysed additional changes in the LC caused by stellar spots and pulsations in the next two sections (\ref{spot} and \ref{puls}). Finally, the paper is concluded with a discussion of the results in Sec.~\ref{conc}.

\section{Morphological changes on the LC}
\label{lc}

KIC\,7023917 was observed during the \textit{Kepler} mission in long-cadence mode (with a time resolution of half an hour)\footnote{\textit{Kepler} data can be found in MAST \cite{kepler}: \dataset[10.17909/T9488N]{http://dx.doi.org/10.17909/T9488N}.}. \textit{TESS} satellite observed this target in multiple sectors (S14, S40, S41, and S54) with a time resolution of two minutes\footnote{\textit{TESS} data are available in MAST \cite{tess,tess-ffi}: \dataset[10.17909/t9-nmc8-f686]{http://dx.doi.org/10.17909/t9-nmc8-f686N} and \dataset[10.17909/0cp4-2j79]{http://dx.doi.org/10.17909/0cp4-2j79}.}. We used PDCSAP flux data for \textit{Kepler} data and data from these \textit{TESS} sectors. Observations during five additional sectors are planned for 2024. Only full-frame images (FFI) with our target will be collected during these sectors with the cadence shortened to 200~s. The extracted LCs are similar to those from previous sectors.
%in the two-minute cadence. 
The first two of these additional sectors (S74, S75) are already available. To extract LC from \textit{TESS} FFI, we used package \textsc{Lightkurve} \citep{Lightkurve}.

\textit{Kepler} data were used mainly for the analysis of the general shape of the LC and to study its long-term changes (this Section and Sec.~\ref{spot}). However, this type of data is unusable for studying short-period signals (such as pulsations). We used observations from \textit{TESS} for this purpose.

\begin{figure}[h]
\centering
\includegraphics[width=\columnwidth]{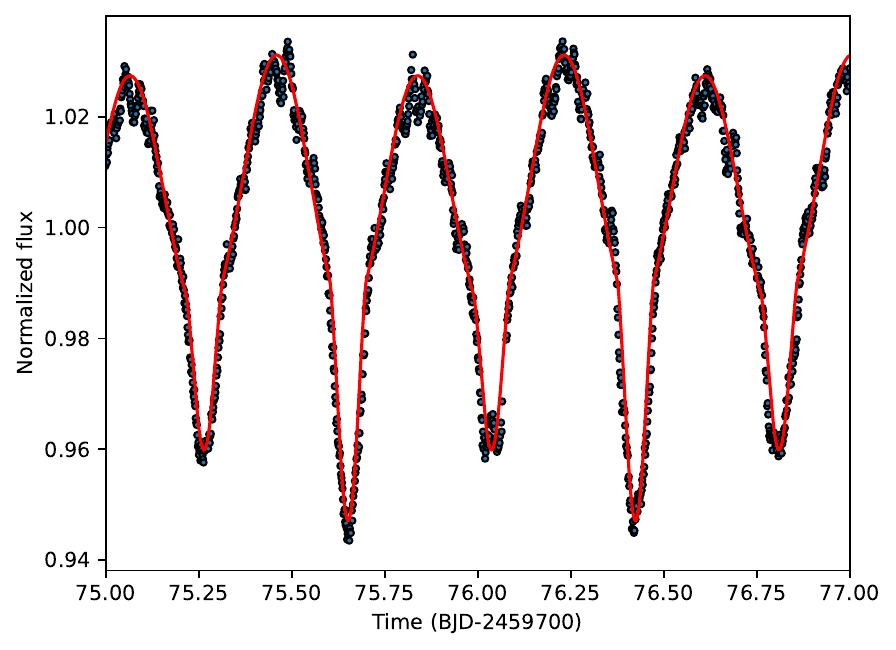}
\caption{Example of the LC of KIC 7023917 obtained using \textit{TESS} satellite . The solid line represents the LC model described in Sec.~\ref{model} which includes the stellar spot (Sec.~\ref{spot}).}
\label{fig:lc}
\end{figure}

Figure~\ref{fig:lc} shows a part of the \textit{TESS} LC. Based on its shape, this EB can be classified as $\beta$~Lyrae type. This classification was also suggested by \cite{Shi2022} and is in good agreement with the analytical model presented later (Sec.~\ref{model}). Two additional phenomena are observed in the LC. The heights of the maxima are unequal, and this feature is also present in the \textit{Kepler} data. We examine this O'Connell effect in Sec.~\ref{oconnell}. The larger but quite regular scatter of data (mainly visible close to the maxima) could be caused by stellar pulsations. These variations are too fast to be detectable in the \textit{Kepler} data. Additionally, the LC is also affected by changes in the determined times of the minima. In this section, we study changes in the O-C diagram (as a visualization of the difference between observed minima times and calculated/predicted ones) and the O'Connell effect and attempt to combine these morphological changes of the LC.

\subsection{O-C diagram}
\label{oc}

First, we used all observations from the \textit{Kepler} and \textit{TESS} missions to obtain a precise linear ephemeris. In addition, we include four minima observed by \textit{SuperWASP} project \citep{wasp,wasp-archive} to extend the time range of the observations. We detected individual primary and secondary minima and determined their mid-times by approximating their shape by the Gaussian function and following fitting using the Monte Carlo method. Obtained times were fitted using our software \textsc{OCFit} \citep{Gajdos2019} with the result of the linear ephemeris for the primary minima in a form:
\begin{equation}
\label{eq:eph}
%T_{\rm I} = {\rm BJD}~2\,454\,953.889535 (46) + 0.772842086 (10) \times E\,,
T_{\rm I} = {\rm BJD}~2\,454\,953.8899455 (80) + 0.77284224296 (79) \times E\,,
\end{equation}
where $E$ is the epoch of the observation and uncertainties are given in parentheses.

\begin{figure}[t]
\centering
\includegraphics[width=\columnwidth]{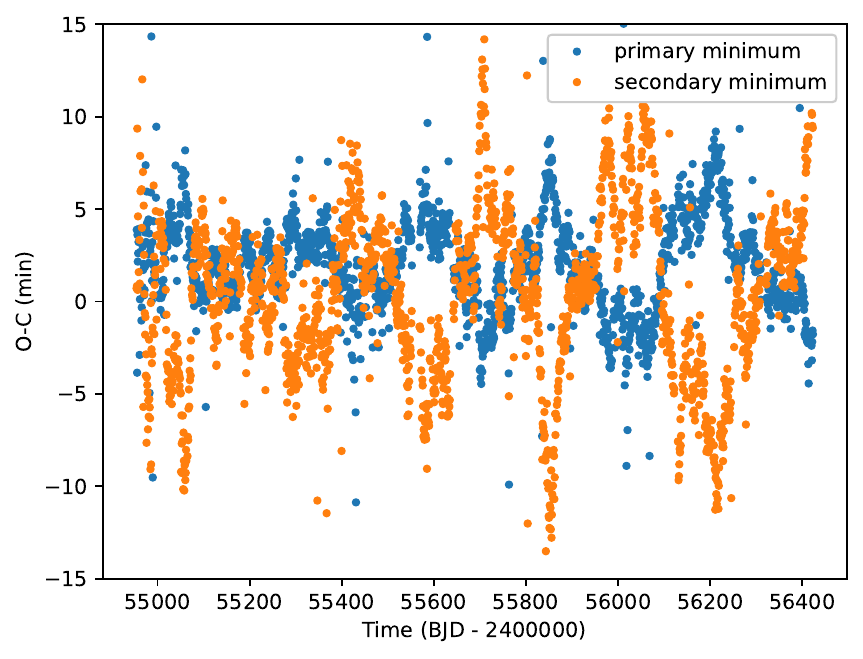}
\includegraphics[width=\columnwidth]{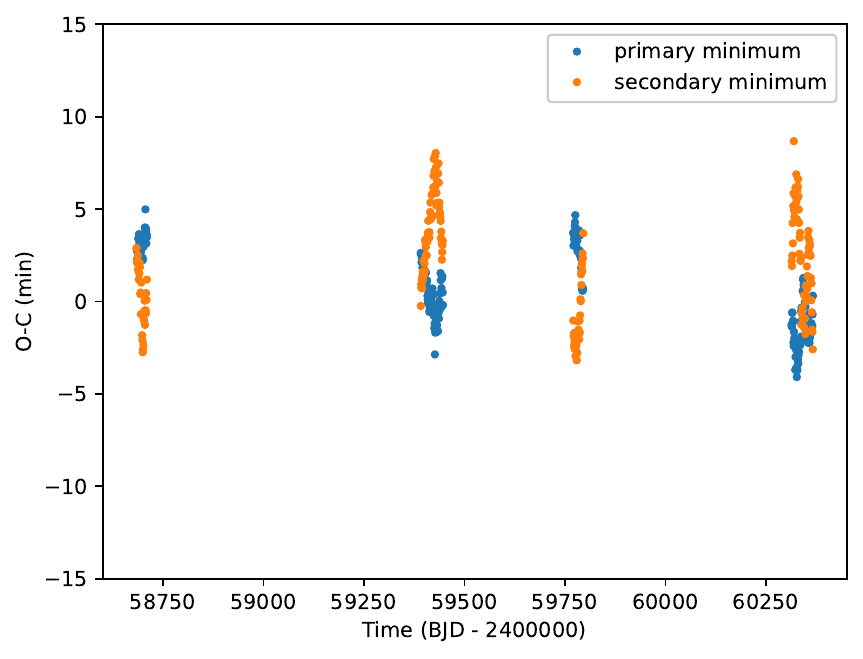}
\caption{O-C diagrams based on data from \textit{Kepler} (top) and \textit{TESS} missions (bottom).}
\label{fig:oc}
\end{figure}

Using the improved ephemeris, we constructed the O-C diagram displayed in Fig.~\ref{fig:oc}. The times of the primary and secondary minima show quasi-periodic anti-correlated variations. Their amplitude is on the level of five minutes with a time scale in the range of 200 -- 300 days. Data from \textit{TESS} has a similar behaviour, perhaps with a slightly lower scatter than the \textit{Kepler} ones. However, their time coverage (with gaps of hundreds of days) is insufficient for precise analysis.

%\begin{figure}[h]
%\centering
%\includegraphics[width=\columnwidth]{min-separation1}
%\caption{Phase separation between following secondary and primary minima.}
%\label{fig:oc-phase}
%\end{figure}

Similar changes in the O-C diagram can be considered as a result of apsidal motion. A similar apsidal motion would be extremely fast, with a periastron advance rate of $\dot{\omega}$ $\sim$0.9~deg/cycle. Moreover, we can expect that an EB with such a short orbital period will have an exactly circular orbit that excludes any apsidal motion. However, an anti-correlated variation in the O-C diagram could also be caused by different effects. \cite{Balaji2015} show that stellar spots could generate such changes. 

%Figure~\ref{fig:oc-phase} shows the phase separation between the secondary and primary minima in the same orbital epoch. This figure is equivalent to the presented O-C diagram and is used in the following analysis. A phase separation of 0.5, means no change in the time of any minima (i.e. O-C is zero).

\subsection{O'Connell effect}
\label{oconnell}

To study the O'Connell effect, we determined the heights of all maxima on the LC using a similar method as in the case of minima times. We considered this height to be the difference between the maximum and mean LC levels. Additionally, to distinguish the types of maxima, we defined the primary maximum as the maximum following the primary minimum (i.e., with a photometric phase of approximately 0.25) and similarly for the secondary one (phase close to 0.75).

\begin{figure}[t]
\centering
\includegraphics[width=\columnwidth]{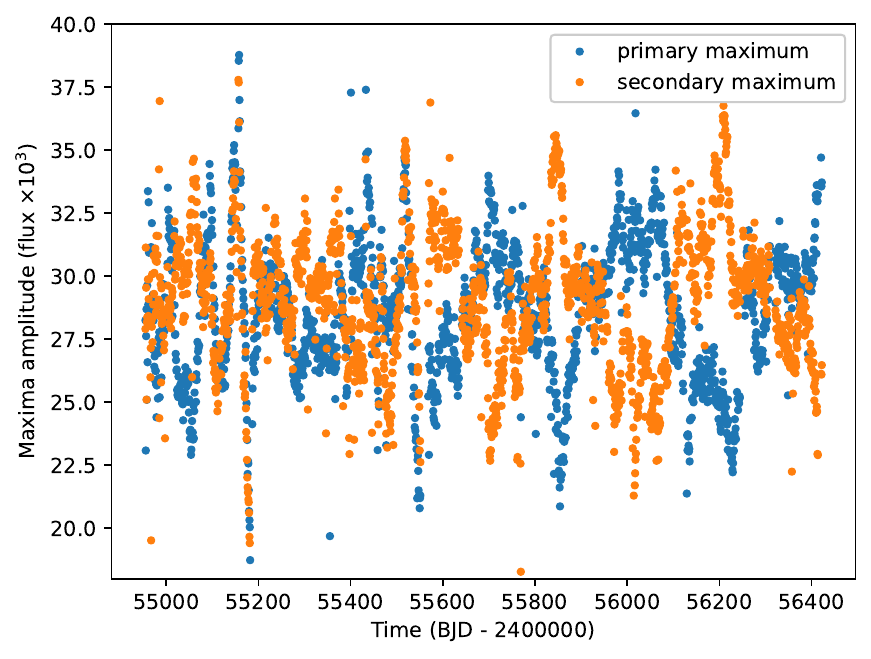}
\includegraphics[width=\columnwidth]{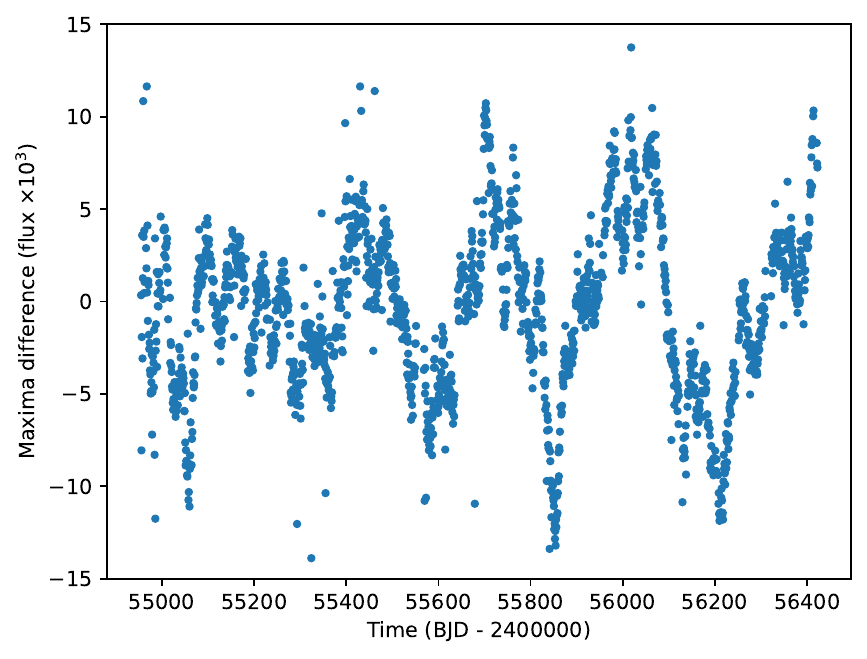}
\caption{Amplitudes of maxima (\textit{top}) and differences between heights of following maxima (\textit{bottom}).}
\label{fig:oconnell}
\end{figure}

Measured maxima heights are in a range from 20 to 40~mmag (with a mean level of about 29~mmag; Fig.~\ref{fig:oconnell}). There is a significant anti-correlation between the amplitudes of the primary and secondary maxima, mainly in the second part of the \textit{Kepler} observation period. We can observe very similar variations in the O-C diagram presented above (Sec.~\ref{oc} and Fig.~\ref{fig:oc}). If we look at the O'Connell effect as the difference between the heights of the maxima, we obtain its quasi-periodic evolution over time (Fig.~\ref{fig:oconnell}). The mean value is close to zero and the values change around it symmetrically, which means that, on average, both maxima have equal heights, and they only change over time. 

%\begin{figure}[t]
%\centering
%%\includegraphics[width=\columnwidth]{offset1}
%\includegraphics[width=\columnwidth]{separation1}
%\caption{Phase separation between following maxima.}  %Phase offset of maxima (\textit{top}) from phase 0.25 (primary maximum) or 0.75 (secondary one). 
%\label{fig:oconnell-phase}
%\end{figure}

%We also determined the phase offset (phase difference between the observed maximum and the phase 0.25 or 0.75 for the secondary one) for all maxima, which could be considered as some O-C diagram for them. We found that primary maxima always occur earlier than expected (phase smaller than 0.25); otherwise, secondary ones are late with phases larger than 0.75. If we recalculated these phase offsets to times, we obtained values for these O-Cs in the range of 5 to 8 min. This separation of phase offsets of maxima is also visible in times when the O'Connell effect is not very strong and the heights of the maxima are nearly identical. However, during the "strong O'Connell effect", the changes in phase offsets are more dramatic. As expected, the phase separation is greater than 0.5, and the mean value is approximately 0.51. The time evolution of this separation differs slightly from that of the difference in maximum heights, and we did not observe any significant correlation between them.

\subsection{Correlations}

At first sight, the behaviour of the minima times in the O-C diagram (Fig.~\ref{fig:oc}) and maxima heights (Fig.~\ref{fig:oconnell}) are similar. Therefore, we consider that both are the result of one common effect. The deformation of the LC decreases the maximum height and causes a shift in the determined minimum time.

%\begin{figure}[t]
%\centering
%\includegraphics[width=\columnwidth]{kepler_oc-amp1}
%%\includegraphics[width=\columnwidth]{kepler_oc-amp2}
%\caption{Correlation between O-C for minima and amplitude of following maxima.% (\textit{top}). Correlation between O-C for minima and amplitude of previous maxima (\textit{bottom}). 
%Regression slope $k$ and correlation coefficient $r$ are given for both types of minima (distinguished by colour). Solid lines represent model correlation obtained by linear regression.}
%\label{fig:corr}
%\end{figure}

%Because the changes of maxima amplitudes and times of minima for primary and secondary minima/maxima show the same characteristic (only in the opposite directions), we could study the proposed correlation in two ways, depending on the pairing types of minima and maxima. 
If we look at the O-C values for the minima and heights of the following maxima, they are anti-correlated.  %(i.e., join primary minimum with primary maximum -- phase 0.25)
%Similarly, we can study the correlation between minima and previous maxima (that means, for primary minimum use secondary maximum -- phase 0.75). In such cases, the studied changes correlate with each other. 
%In both cases, the correlation was very strong. 
The correlation coefficient is greater than 97\%. The relations for the primary and secondary minima differ slightly. In general, the same change in the maximum amplitude causes a smaller shift in the time of the primary minimum than in that of the secondary minimum. This could result from deeper primary minima, and therefore, are more resistant to LC deformation. The value of O-C equal to zero (i.e. no-shift of minimum time) corresponds to a maxima amplitude of about 30~mmag which is nearly the same for both kinds of maxima. This value can be considered the height of the maxima on the non-deformed LC.

%\begin{figure}[h]
%\centering
%\includegraphics[width=\columnwidth]{kepler_sep-diff1.pdf}
%\caption{Correlation between phase separation of minima and the difference between heights of maxima during the same epoch. Regression slope $k$ and correlation coefficient $r$ together with a model correlation (solid line) are given.}
%\label{fig:corr-phase}
%\end{figure}

We observed a similarly strong correlation between the phase separation of the minima and the difference in the maxima heights. The correlation coefficient is greater than 98\% in this case. Equal maxima amplitudes are associated with a minima phase separation of 0.5. Therefore, we can conclude that the non-deformed LC is the LC of EB with a circular orbit, which is also expected from a short orbital period with a strong circularization mechanism. In the case of an eccentric orbit, the secondary minima would also show some time shift without any LC deformation, and minima would occur in a phase different from 0.5.

\section{Spectroscopy}
\label{spec}

Our spectroscopic observations of KIC\,7023917 were performed using an Ond\v{r}ejov Echelle Spectrograph (OES) installed at the 2-m Perek telescope \citep{Kabath2020} at the Ond\v{r}ejov Observatory (14\degr~46\arcmin~52\arcsec~E, 49\degr~54\arcmin~55\arcsec~N, 525~m a.s.l.). OES is a fibre-fed high-resolution spectrograph with resolving power $R = 50 000$ sensitive between 3600 and 9500~\AA\ which uses ThAr spectra for wavelength calibration.

A total of  18 spectra were obtained in September 2023 (see Tab.~\ref{tab:rv}). However, their quality was strongly affected by the local weather conditions in Ond\v{r}ejov.
%Depending on the level of seeing, humidity, and transparency, the signal-to-noise ratio (SNR) varies from 10 to 50. 
Moreover, we had to use an exposure time of only half an hour (1800~s) to avoid blurring lines caused by orbital motion with a short period of approximately 18~h. Such a short exposure time is at the limit of usability of used equipment for the target with a brightness of only 10~mag. 
%Finally, the primary component of this EB is a star with estimated spectral type A, with only a few spectral lines usable for further analysis.

We could only detect the spectral line of one component (the primary one). The secondary component is invisible in our spectra because of the large luminosity ratio between both stars and their low quality. The obtained spectra were also unusable for fitting individual spectral lines and determining the star's basic physical parameters. 
%We tried to align all obtained spectra according to determined radial velocities (Sec.~\ref{rv}) and combine them to improve SNR. We modelled this combined spectrum in \textsc{iSpec} \citep{Blanco2014,Blanco2019}. The determined value of $v\sin i$ is about 100 km/s which gives a rotational period of nearly 26.5 hours (determined using stellar radius obtained in Sec.~\ref{model}).

\subsection{Radial velocities}
\label{rv}
We measured the radial velocity (RV) of the primary component of the KIC\,7023917 and its uncertainties using the cross-correlation function in package \textsc{iSpec} \citep{Blanco2014,Blanco2019}. We selected only the spectral region around the magnesium triplet (5100~--~5200~\AA). As a template, we used one of our spectra with a higher SNR, obtained in 5~September (JD~2\,460\,193.2737). The SNR was determined in \textsc{iSpec} from the observed flux.

\begin{table}[h]
\caption{Measured relative radial velocities of the primary component.}             
\label{tab:rv}      
\centering                        
{\scriptsize \begin{tabular}{ccccc}
	\hline\hline
	Julian date  & Phase & RV (km/s) & Error (km/s) & SNR \\ \hline
	2460192.3829 & 0.219 &   -0.45   &     3.23     & 31  \\
	2460192.4200 & 0.268 &   -1.07   &     3.31     & 30  \\
	2460193.2737 & 0.372 & \,\,0.00  &     0.75     & 42  \\
	2460193.3824 & 0.513 & \,\,0.06  &     0.91     & 30  \\
	2460197.2960 & 0.577 &   18.08   &     3.18     & 22  \\
	2460197.4256 & 0.744 &   39.69   &     4.43     & 35  \\
	2460199.2954 & 0.164 & \,\,6.30  &     4.46     & 34  \\
	2460199.4479 & 0.361 &   -8.34   &     4.47     & 23  \\
	2460200.3115 & 0.478 &   10.78   &     3.87     & 27  \\
	2460200.4504 & 0.658 &   34.31   &     3.61     & 30  \\
	2460202.2644 & 0.005 &   37.19   &     3.17     & 24  \\
	2460202.3507 & 0.117 &   22.50   &     3.21     & 21  \\
	2460203.3084 & 0.356 &   -1.02   &     4.27     & 42  \\
	2460203.5517 & 0.671 &   33.20   &     2.17     & 20  \\
	2460204.4456 & 0.828 &   33.10   &     7.29     & 31  \\
	2460204.5139 & 0.916 &   34.02   &     5.68     & 33  \\
%	2460205.3086 & 0.944 & \,\,8.63  &     9.56     & 46  \\
	2460205.3820 & 0.039 &   32.64   &     4.51     & 28  \\
	2460207.3120 & 0.537 &   11.20   &     4.65     & 23
\end{tabular}}
\end{table}

\begin{figure}[h]
\centering
\includegraphics[width=\columnwidth]{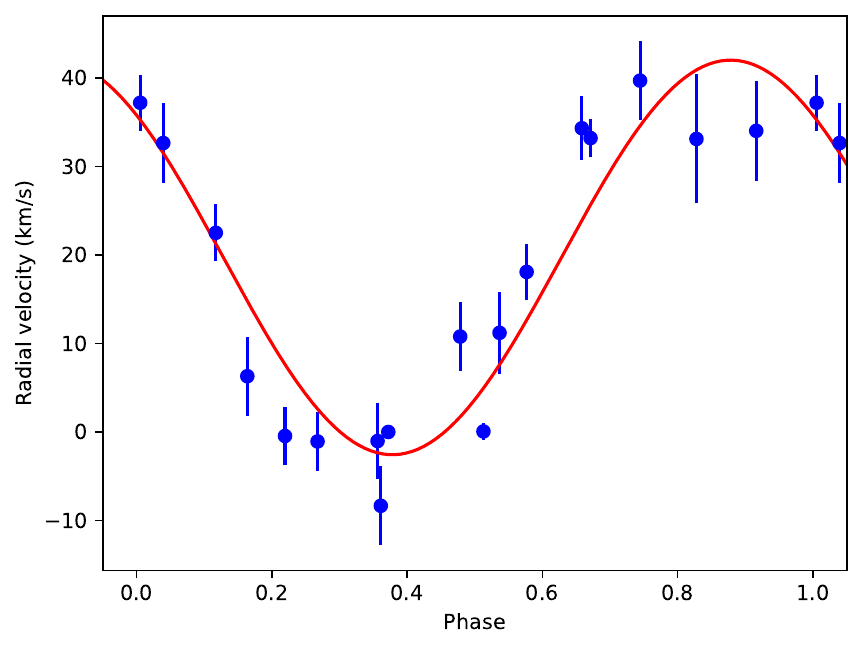}
\caption{Radial velocity measurements of the primary component of  KIC\,7023917  together with a synthetic model (solid line).}
\label{fig:rv}
\end{figure}

The resulting relative RV are presented in Tab.~\ref{tab:rv} and are plotted in Fig.~\ref{fig:rv}. We obtained a similar RV in the spectral region close to the Ballmer lines H$_\gamma$ and H$_\delta$ (4050~--~4450~\AA) used to verify our results. The uncertainties of two RV measured from spectra taken in 5~September (JD~2\,460\,193) are significantly smaller and are not very representative. This is a result of using the first of them as a template. The second one shows also minimal mutual shift (determined RV is only 60~m/s). Other RV uncertainties vary around 4~km/s and are affected by the observing conditions (e.g. seeing, clouds or quality of tracking) and slightly by the difference in RV for the template and analysed spectrum.

%\begin{table}[h]
%\caption{Parameters of  KIC\,7023917  determined from RV modelling.}             
%\label{tab:rv-param}      
%\centering                        
%\begin{tabular}{c|c}
%	\hline\hline
%	    $K_1$ (km/s)     &   $22.28 \pm 2.71$    \\
%	        $e$          &       0 (fixed)        \\
%	   $\omega$ (deg)    &    $43.85 \pm 6.42$    \\
%	$f (M_2)$ (M$_\sun$) & $0.00093 \pm 0.00033 $
%\end{tabular}
%\end{table}

We modelled the RV curve using the standard model of \cite{Irwin1952} with a fixed ephemeris determined from the LC and O-C analysis (Eq.~\ref{eq:eph}). The orbital eccentricity was fixed at 0 based on LC analysis. %However, the shape of the RV curve does not show any significant effect of elliptical orbit.
%The orbit is nearly circular with an eccentricity of less than 0.1 which we already expected from previous analysis. 
The semi-amplitude of the RV is $22.28 \pm 2.71$~km/s. Considering this value, we can calculate the mass function of the secondary component to $0.00093 \pm 0.00033$~M$_\sun$.

Using the inclination value determined by the LC fitting (see Sec.~\ref{model} and Tab.~\ref{tab:all-params}) and the mass of the primary component estimated according to its temperature \citep[$M_1=1.8$~M$_\sun$;][]{Pecaut2013}, we obtained the mass ratio $q \approx 0.10$. This value is consistent with the LC analysis.

\section{Ground-based photometry}
\label{phot}

KIC\,7023917 was observed at the Kolonica Astronomical Observatory of the P. J. \v{S}af\'{a}rik University in Ko\v{s}ice (22\degr~16\arcmin~26\arcsec~E, 48\degr~56\arcmin~05\arcsec~N, 442~m a.s.l.). We used the Maksutov-Newtonian telescope with a diameter of 152~mm and a focal ratio of f/4.8, together with a G2-8300 CCD camera equipped with Sloan $g'r'$ photometric filters. 

\begin{table}[h]
\caption{Number of obtained ground-based photometric data during individual nights.}             
\label{tab:obs}      
\centering                        
{\scriptsize \begin{tabular}{cccc}
	\hline\hline
	   Date    &    Time UT     &    Phase     & No. obs. ($g'/r'$) \\ \hline
	2023-Jul-11 & 21:05 -- 00:00 & 0.05 -- 0.21 &      88 / 87       \\
	2023-Jul-14 & 21:35 -- 01:25 & 0.96 -- 0.17 &     102 / 118      \\
	2023-Jul-15 & 21:45 -- 00:40 & 0.26 -- 0.42 &      65 / 93       \\
	2023-Jul-23 & 20:20 -- 23:45 & 0.54 -- 0.72 &      101 / 94      \\
%	2023-Jul-28 & 21:25 -- 21:45 & 0.06 -- 0.08 &       6 / 2        \\
	2023-Jul-31 & 22:10 -- 00:25 & 0.98 -- 0.11 &      69 / 69       \\
%	2023-Aug-03 & 20:00 -- 22:45 & 0.76 -- 0.77 &       2 / 1        \\
	2023-Aug-05 & 19:35 -- 00:37 & 0.32 -- 0.53 &      52 / 52       \\
	2023-Sep-16 & 18:10 -- 01:40 & 0.58 -- 0.89 &     140 / 132      \\
	2023-Sep-17 & 18:05 -- 01:00 & 0.87 -- 0.25 &     178 / 183      \\
	2023-Sep-25 & 19:15 -- 03:30 & 0.29 -- 0.64 &     175 / 155      \\
	2023-Oct-13 & 20:25 -- 03:35 & 0.66 -- 0.83 &      63 / 52
\end{tabular}
\tablecomments{Only observations used in the following analysis are listed.}}
\end{table}

\begin{figure}[h]
\centering
\includegraphics[width=\columnwidth]{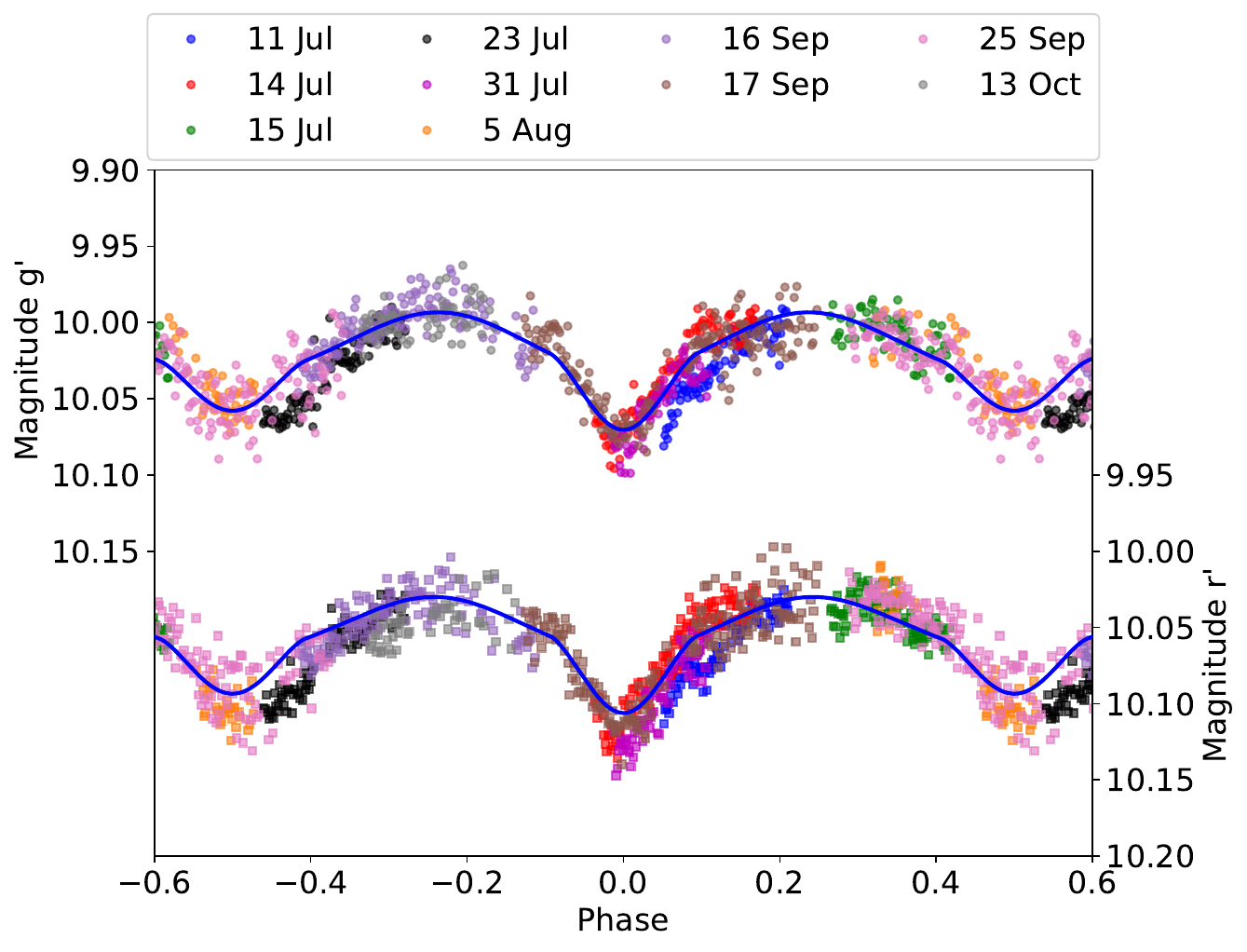}
\caption{Collected phase-folded LC in $g'$ and $r'$ filters by dates of observations (see details in Tab.~\ref{tab:obs}). Solid blue curves are synthetic models based on the parameters from Sec.~\ref{model}.}
\label{fig:lc-gr}
\end{figure}

Our observations during 10 nights in summer and autumn 2023 cover all photometric phases (see Tab.~\ref{tab:obs} and Fig.~\ref{fig:lc-gr}). 
%We observed using $g'$, $r'$ and $i'$ filters. However, the data obtained in $i'$ are unusable because of the large scatter caused by local weather conditions. 
We performed aperture photometry by calculating the differential magnitudes according to nearby comparison stars with similar magnitudes and spectral types. The obtained magnitudes were recalculated to standard ones using standard APASS magnitudes of comparison stars \citep{apass}. The interstellar reddening measured by \textit{Gaia} mission with value $E (B-V) = 0.028~mag$ was applied in the following calculations.

%\begin{table}[h]
%\caption{Estimated colour indices and temperature from multi-colour photometry. \textbf{ZRUSIT!}}             
%\label{tab:color}      
%\centering                        
%\begin{tabular}{c|c}
%	\hline\hline
%	$g' - r'$ & $-0.0378 \pm 0.0095$ \\
%	  $B-V$   & $0.1972 \pm 0.0128$  \\
%	  $V-R$   & $0.0704 \pm 0.0087$  \\
%	 $T$ (K)  &    $7587 \pm 81$
%\end{tabular}
%\end{table}

We can assume that any variations caused by the O'Connell effect and changes in the O-C diagram are eliminated on phase-folded LC from our observations covering a long time interval. In addition, these effects would probably be indistinguishable in our ground-based photometric data. We fitted the folded LC with a model (for details, see Sec.~\ref{model}). The obtained results are comparable to those for the data from \textit{Kepler} and \textit{TESS}, but with significantly larger uncertainties (listed in Tab.~\ref{tab:all-params}).

The main goal of our ground-based multicolour photometry was to estimate the colour indices and temperature of the primary star. We have used equations of Robert Lupton\footnote{\url{https://www.sdss3.org/dr10/algorithms/sdssUBVRITransform.php}} to transform measured magnitudes in Sloan filters $g'$ and $r'$ to the Bessel ones $B$, $V$ and $R$. The calculated colour index $g'-r'$ is $-0.0378 \pm 0.0095$. Using the colour index $B-V = 0.1972 \pm 0.0128$, we estimated the effective temperature to $7587 \pm 81$~K \citep{Sekiguchi2000}, which is approximately 100~K above the value given by the \textit{Gaia} mission. The quality and scatter of our observations probably cause this difference. The presence of a secondary component may induce some discrepancies. We can also determine colour index $V-R$ to $0.0704 \pm 0.0087$. However, there is no known relation between effective temperature and this colour index.

\section{LC modelling}
\label{model}

For LC analysis, we used \textsc{ELISa} code \citep{Cokina2021}. This Python package was designed to model EBs, including surface features such as spots. %\textsc{ELISa} utilizes modern approaches to EB modelling with an emphasis on computational speed, while maintaining a sufficient level of precision to process ground-based and space-based observations.

\begin{table*}
\caption{Photometric parameters of the EB obtained from different sources of observations (the first two parts of the table). Absolute parameters (star radii, solar masses and semi-major axis) were calculated using the amplitude of RV $K_1$ and based on \textit{Gaia} distance, independently (the next parts).}           
\label{tab:all-params}      
\centering                        
{\footnotesize \begin{tabular}{c|cccc}
	\hline\hline
	                      &            \textit{Kepler}            &        \textit{TESS}         &        \textit{Kepler+TESS}        &          $g'+r'$           \\ \hline
	         $q$          &  $0.079576_{-0.000043}^{+0.000011}$   &  $0.105_{-0.017}^{+0.011}$   & $0.080287_{-0.000015}^{+0.000040}$ & $0.071_{-0.012}^{+0.013}$  \\
	      $i$ (deg)       &     $58.4531_{-0.0023}^{+0.0043}$     &   $58.59_{-0.21}^{+0.27}$    &   $58.3838_{-0.0038}^{+0.0022}$    &  $58.85_{-0.62}^{+0.83}$   \\
	     $\Omega_1$       &   $2.19737_{-0.000055}^{+0.000077}$   &  $2.251_{-0.029}^{+0.030}$   & $2.199537_{-0.000058}^{+0.000121}$ & $2.095_{-0.081}^{+0.085}$  \\
	     $\Omega_2$       &  $1.894005_{-0.000136}^{+0.000035}$   &  $2.004_{-0.079}^{+0.064}$   & $1.896518_{-0.000048}^{+0.000123}$ & $1.881_{-0.043}^{+0.048}$  \\
	     $T_1$  (K)       &                                        \multicolumn{4}{c}{\tablenotemark{a} $7461_{-15}^{+16}$}                                         \\
	     $T_2$   (K)      &       $6397.97_{-0.44}^{+0.45}$       &      $6411_{-22}^{+22}$      &     $6393.86_{-0.43}^{+0.43}$      &     $6220_{-70}^{+69}$     \\ \hline
	     $\Omega_C$       &  $1.893949_{-0.000143}^{+0.000036}$   &  $1.975_{-0.053}^{+0.034}$   &  $1.8963_{-0.000050}^{+0.000134}$  & $1.864_{-0.042}^{+0.045}$  \\
	     $R_1^{eq}$       &  $0.493655_{-0.000013}^{+0.000013}$   & $0.486_{-0.0056}^{+0.0061}$  & $0.493288_{-0.000021}^{+0.000011}$ & $0.520_{-0.021}^{+0.022}$  \\
	     $R_2^{eq}$       & $0.1919252_{-0.0000344}^{+0.0000081}$ & $0.1952_{-0.0055}^{+0.0034}$ & $0.192339_{-0.000012}^{+0.000035}$ & $0.179_{-0.015}^{+0.011}$  \\ %\hline
	   %   fit $R^2$       &               0.997448                &           0.995439           &              0.997453              &          0.936699          \\
	%   fit $\chi_r^2$     &                 1.20                  &             1.35             &                1.30                &            1.03
	\hline\hline
		                      &                                          \multicolumn{4}{c}{from RV ($K_1$)}                                          \\
		   $a$ (R$_\sun$)     &  $5.419_{-0.660}^{+0.660}$   & $4.189_{-0.563}^{+0.570}$  &  $5.379_{-0.691}^{+0.691}$   & $6.000_{-0.867}^{+0.898}$  \\
		$R_1^{eq}$ (R$_\sun$) &  $2.818_{-0.360}^{+0.360}$   & $2.053_{-0.275}^{+0.278}$  &  $2.614_{-0.341}^{+0.341}$   & $2.916_{-0.469}^{+0.484}$  \\
		$R_2^{eq}$ (R$_\sun$) &  $0.970_{-0.143}^{+0.143}$   & $0.819_{-0.112}^{+0.114}$  &  $1.050_{-0.133}^{+0.133}$   & $1.171_{-0.183}^{+0.187}$  \\
		  $M_1$ (M$_\sun$)    &  $3.309_{-1.209}^{+1.209}$   & $1.503_{-0.693}^{+0.701}$  &  $3.234_{-1.428}^{+1.428}$   & $4.528_{-2.319}^{+2.326}$  \\
		  $M_2$ (M$_\sun$)    &  $0.263_{-0.096}^{+0.096}$   & $0.158_{-0.073}^{+0.074}$  &  $0.260_{-0.115}^{+0.115}$   & $0.321_{-0.165}^{+0.165}$  \\ \hline
		                      &                                  \multicolumn{4}{c}{based on \textit{Gaia} distance}                                  \\
		   $a$ (R$_\sun$)     &  $4.454_{-0.020}^{+0.019}$   & $4.510_{-0.449}^{+0.293}$  &  $4.458_{-0.019}^{+0.020}$   & $4.292_{-0.951}^{+0.817}$  \\
		$R_1^{eq}$ (R$_\sun$) &  $2.181_{-0.009}^{+0.009}$   & $2.175_{-0.012}^{+0.010}$  &  $2.180_{-0.009}^{+0.009}$   & $2.206_{-0.014}^{+0.013}$  \\
		$R_2^{eq}$ (R$_\sun$) & $0.8620_{-0.0002}^{+0.0002}$ & $0.885_{-0.036}^{+0.023}$  & $0.8646_{-0.0002}^{+0.0002}$ & $0.777_{-0.070}^{+0.062}$  \\
		  $M_1$ (M$_\sun$)    &  $1.839_{-0.024}^{+0.024}$   & $1.861_{-0.557}^{+0.363}$  &  $1.842_{-0.024}^{+0.024}$   & $1.659_{-1.103}^{+0.948}$  \\
		  $M_2$ (M$_\sun$)    &  $0.146_{-0.002}^{+0.002}$   & $0.199_{-0.093}^{+0.061}$  &  $0.148_{-0.002}^{+0.002}$   & $0.118_{-0.098}^{+0.089}$  \\
		  $L_1$ (L$_\sun$)    &  $13.275_{-0.002}^{+0.002}$  & $13.207_{-0.092}^{+0.058}$ &  $13.272_{-0.002}^{+0.002}$  & $13.584_{-0.131}^{+0.116}$ \\
		  $L_2$ (L$_\sun$)    & $1.1218_{-0.0003}^{+0.0003}$ & $1.191_{-0.095}^{+0.060}$  & $1.1257_{-0.0004}^{+0.0004}$ & $0.815_{-0.141}^{+0.125}$  \\
		                      &                               \multicolumn{4}{c}{masses from \textit{Gaia} + $f(M_2)$}                                \\
		  $M_1$ (M$_\sun$)    &  $1.806_{-0.048}^{+0.047}$   & $1.865_{-0.660}^{+0.459}$  &  $1.799_{-0.047}^{+0.048}$   & $1.600_{-1.262}^{+1.085}$  \\
		  $M_2$ (M$_\sun$)    &  $0.180_{-0.022}^{+0.022}$   & $0.195_{-0.045}^{+0.035}$  &  $0.191_{-0.022}^{+0.022}$   & $0.176_{-0.081}^{+0.070}$  \\
		         $q$          &  $0.100_{-0.014}^{+0.014}$   & $0.105_{-0.061}^{+0.044}$  &  $0.106_{-0.015}^{+0.015}$   & $0.110_{-0.137}^{+0.119}$  \\ 
\end{tabular}}
\tablenotetext{a}{Fixed value adopted from \textit{Gaia} DR3 database.}
\end{table*}

The Levenberg-Marquardt least-squares method was used to find the initial approximate solutions for different types of input datasets: \textit{Kepler} data, \textit{TESS} data,  \textit{Kepler} \& \textit{TESS} data, and ground-based observations ($g'$ \& $r'$). The final set of parameters with their uncertainties is a result of Markov chain Monte Carlo (MCMC) sampling. The model contains five free parameters: orbital inclination $i$, photometric mass ratio $q$, surface potentials of both components $\Omega_1$ and $\Omega_2$ and effective temperature of the secondary component $T_2$. The temperature of the primary component $T_1$ was adopted from the \textit{Gaia} DR3 database and fixed during the fitting routine. We did not consider stellar spots at this stage. We used all available data from \textit{Kepler} and \textit{TESS} missions covering a few years of observation, which smoothened out any spot effect (such as the O'Connell effect) with significantly shorter periods. 

\begin{figure}[h]
\centering
\includegraphics[width=\columnwidth]{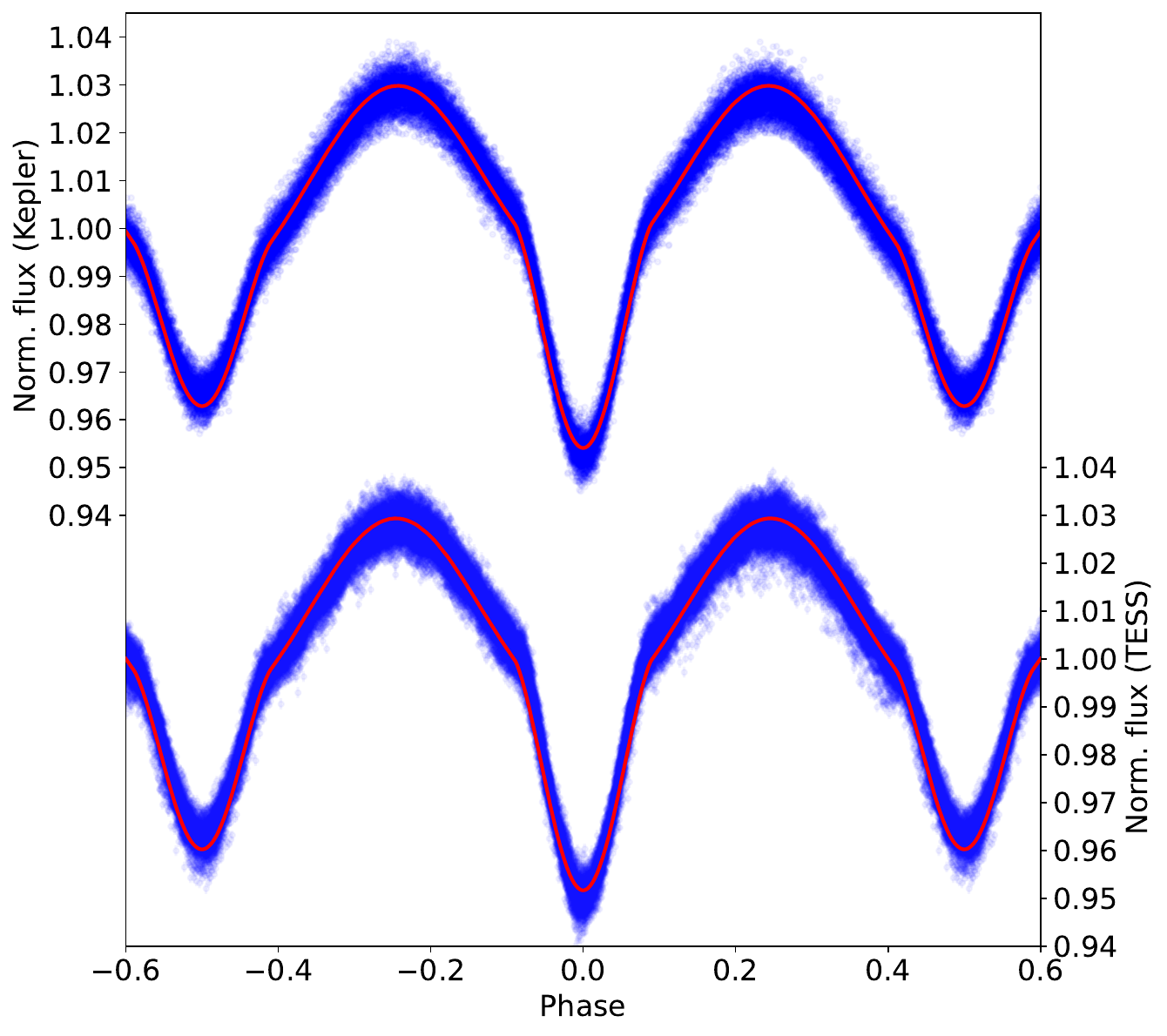}
\caption{Phase-folded LC obtained by \textit{Kepler} and \textit{TESS} mission with model (solid line).}
\label{fig:model-lc}
\end{figure}

\begin{figure}[h]
\centering
\includegraphics[width=\columnwidth]{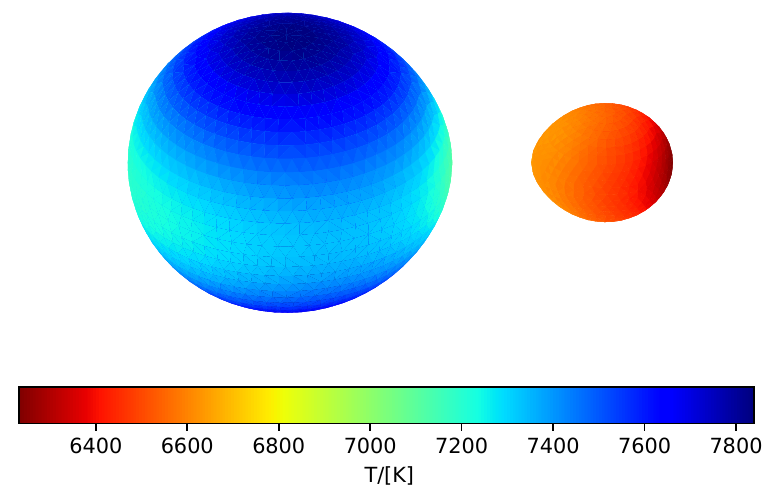}
\includegraphics[width=\columnwidth]{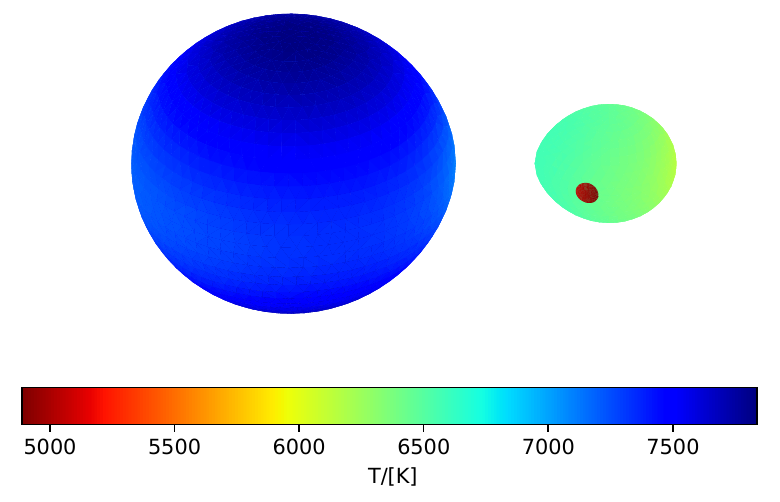}
\caption{A 3D model with the surface temperature distributions. The model is calculated for the parameters from Tab.~\ref{tab:all-params} based on \textit{TESS} observations (\textit{top}). The model with predicted star spot (see Sec.~\ref{spot}) is shown on \textit{bottom} panel.}
\label{fig:model}
\end{figure}

The resulting parameters are listed in the first part of Tab.~\ref{tab:all-params}. The main difference between the values for the \textit{Kepler} and \textit{TESS} data is in the mass ratio, which affects the Roche geometry of the system and, therefore, partially affects the surface potentials of the components $\Omega_{1,2}$. Some additional parameters (critical potential $\Omega_C$ and equivalent radii of the stars $R^{eq}_{1,2}$ in units of the semi-major axis) were derived and are listed in the next part of this table. Model LC with the data from \textit{Kepler} and \textit{TESS} missions are displayed in Fig.~\ref{fig:model-lc}. 

Moreover, the errors of all parameters determined for \textit{Kepler} data are too small to be realistic. We used the Monte Carlo method to estimate errors of fitted parameters. However, this method is susceptible to realistic errors of input data. For \textit{Kepler} data, these input uncertainness are significantly underestimated (maybe due to used binning). Because the following analysis (see below) shows that parameters from the \textit{TESS} data better described the observed system, we put these errors for \textit{Kepler} data here only for formal reasons without any additional re-scaling. The obtained uncertainties for \textit{TESS} data look more reasonable.
%The corresponding corner plot is shown in Fig.~\ref{fig:corner}.

In summary, the orbital inclination of the system is approximately 58.5\degr, and the effective temperature of the secondary component is 6400~K, which corresponds to spectral type F6 or F7. The obtained mass ratio is small (0.07~--~0.10) which is close to the value determined from our RV measurements (Sec.~\ref{rv}). The radius of the primary component is nearly half of the mutual distance between the components. The secondary star almost fills its Roche lobe, as illustrated in Fig.~\ref{fig:model}. In any case, the system should be classified as either detached or semi-detached.

The low mass ratio is unusual for EB, consisting of these spectral-type stars (A7 and F6). The solution could have been in the past mass transfer from the secondary component to the primary one, which could also be indicated by the current Roche geometry of the system with one of the lobes almost filled with the secondary star. The filling factor of the secondary component is greater than 98.5\%. The mass transfer probably stopped or is at a minimal level. Otherwise, we would detect it as a parabolic trend in the O-C diagram. Hypothetical mass transfer invisible in the O-C diagram should be lower than $10^{-11}$~M$_\sun$/yr.

To decide between the parameters determined from different types of datasets, we used the calculated absolute parameters of the system. There are two possible options for obtaining them. We can use the amplitude of RV $K_1$ together with photometric mass ratio $q$ to calculate the semi-major axis using equation
\begin{equation}
a=\frac{K_1P}{2\pi\sin i}\frac{q+1}{q}\,.
\end{equation}
Subsequently, the radii of the stars could be directly calculated, and their masses were obtained using Kepler's third law. The absolute parameters calculated using this method are listed in the third part of Tab.~\ref{tab:all-params}.

If we know the distance of the EB from the \textit{Gaia} measurements, we can follow the procedure of absolute parameters calculation described in \cite{Kudak2023}. Using this method, we obtained almost all parameters without using the value of the photometric mass ratio. The mass ratio was used only to divide the total system mass into the masses of individual components. However, if we take the mass function $f(M_2)$ determined from RV fitting, we can calculate the mass $M_2$ and then $M_1$ without knowing the mass ratio. Therefore, this method can be used for partially independent estimation of mass ratio $q$. The calculated parameters are listed in the last part of Tab.~\ref{tab:all-params}, which consists of parameters calculated only with photometric parameters and the \textit{Gaia} distance (also using the photometric mass ratio) and separately the masses of the components and their mass ratio obtained from the mass function, as described above.

Comparing the absolute parameters across the different datasets and both calculation methods, we found that the parameters obtained using data from \textit{TESS} are more reliable. They are the most self-consistent and physically acceptable. The parameters using the \textit{Kepler} data differ significantly based on the calculation method used. The mass of the primary component calculated from the RV semi-amplitude is approximately 3.3~M$_\sun$ which is hardly possible for this spectral type and the entire system. The parameters calculated based on distance are closer to those of \textit{TESS}. Moreover, the 'independently' determined mass ratio for all datasets is nearly the same and confirms the photometric value from modelling the \textit{TESS} data. Therefore, we marked the parameters obtained from the \textit{TESS} data as the correct set and used them in the following analysis

We also attempted to confirm the obtained parameters using spectral energy distribution (SED) modelling in the online VOSA tool \citep{Bayo2008}. The temperature difference between the stars is relatively small but the estimated ratio of the stellar luminosities is big. Therefore, we cannot distinguish the contribution of the secondary component to the SED, and only the parameters of the primary one can be determined in this manner. Results are similar to those obtained from the \textit{Gaia} database and our LC fitting. The temperature is estimated to be $7482\pm100$~K, the surface gravity is $4.57\pm0.50$~dex, the metallicity is $-0.63\pm0.30$~dex, and the mass is $1.79\pm0.02$~M$_\sun$.

\section{Stellar spot}
\label{spot}

The possibility of stellar spots on one of the components of KIC\,7023917 has already been reported by \cite{Balaji2015}. This system was marked as one among 414 others, whose O-C diagrams show anti-correlated changes. \cite{Balaji2015} claimed that the stellar spots could explain them. However, they did not study this system in more detail.

\begin{figure}[h]
\centering
\includegraphics[width=\columnwidth]{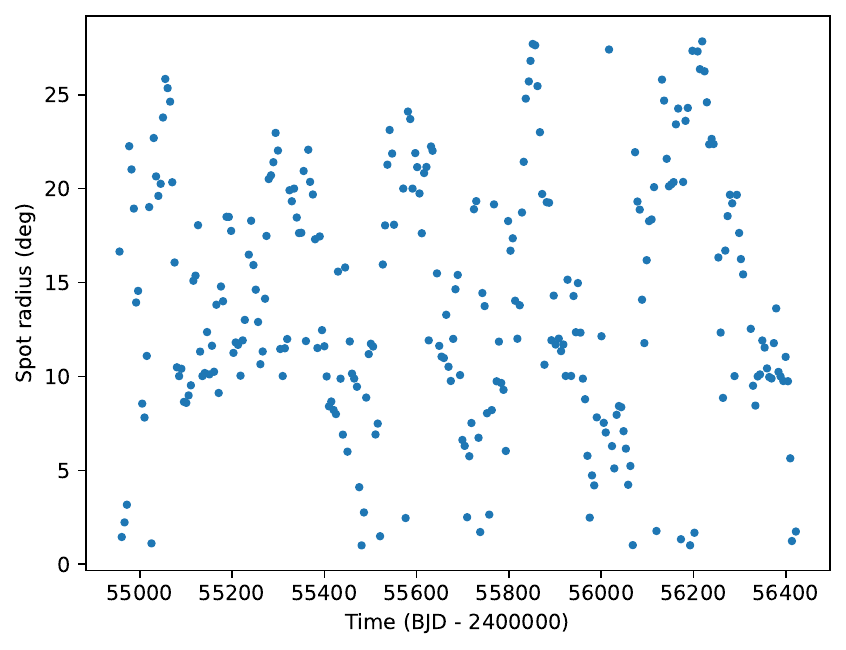}
\caption{Changes of the angular radius of the predicted star spot.}
\label{fig:spot}
\end{figure}

In this section, we attempt to model the LC for  KIC\,7023917  using the parameters obtained in Sec.~\ref{model} and place a stellar spot on the surface of the star. We used \textit{Kepler} data for this purpose. We split the original LC into 5-day-long parts that are sufficiently long to provide enough data points; however, the LC does not change significantly.

%\begin{figure}[h]
%\centering
%\includegraphics[width=\columnwidth]{kepler_spotrad-diff_eq0-1}
%\caption{Correlation between angular radius of the spot and mean difference between heights of LC maxima over the same time interval.}
%\label{fig:spot-corr}
%\end{figure}

We assumed a cold spot located in the secondary component. In principle, it is impossible to distinguish between hot and cold spots or the exact component where the spot is located only on the shape of the observed LC. 
%However, it is more probable that the spot would be on the star of the estimated spectral type of the secondary component (F6~--~F7) than on the hotter primary one \citep[spectral type A7; e.g.][]{Berdyugina2005,Strassmeier2009}. %Moreover, the primary component is pulsating (see Sec.~\ref{puls} for details), which decreases the likelihood of the long-term existence of the spot. 
However, cool spots have bigger brightness contrast compared to chemical, temperature spots in hot stars and are more probable for stars with extended convective envelopes like the secondary component of the estimated spectral type F6~--~F7 than on the hotter primary star \citep[spectral type A7; e.g.][]{Berdyugina2005,Strassmeier2009}.
In addition, cool spots are more common than hot ones.

The individual short parts of the LC were fitted using the parameters determined in Sec.~\ref{model} where a cold spot on the secondary component was added (Fig.~\ref{fig:model}, bottom). Only the parameters of the spot (size and location) and the temperature of the secondary star were fitted. The difference between stellar and starspot temperature was estimated to be about 1600~K \citep{Herbst2021}. Therefore, the spot temperature factor was set to 0.75. We attempted different hypothetical spot configurations. The best results were obtained for a spot near the stellar equator. Its angular size varied around a mean value of approximately 12.5\degr\ (Fig.~\ref{fig:spot}).

The variation of the spot size shows a periodicity of approximately 300 days, and its behaviour is similar to that seen in the O-C diagram (Fig.~\ref{fig:oc}) or at the heights of the maxima (Fig.~\ref{fig:oconnell}). We found a significant correlation between the spot radius and the values of the O'Connell effect (differences in the maximum heights). Growing the stellar spot, the maximum following the primary minimum (primary maximum in our notation) decreases compared with the other one. Therefore, we can explain the observed morphological changes in the LC by the variation in the radius of the cold star spot presented on the secondary component of this EB. These changes in spot size could be interpreted as the evolution of the considered spot or group of star spots similar to observed solar spots. Alternatively, we can explain observed variations in the LC by the changes in the temperature of the fixed-size spot due to the correlation between spot radius and its temperature factor.

\section{Pulsations}
\label{puls}

Pulsations of one component of KIC\,7023917 have already been reported \cite{Shi2022}. They determined only the strongest pulsations with a period of 0.0364023~d (52.419312~min) and an amplitude of 1.6~mmag. This pulsation period was also detected in our following analysis; however, the amplitude was lower by approximately 0.3~mmag. Based on the pulsation period and the relationship between the orbital and pulsation periods \citep{Liakos2012}, they suggested that the pulsations are of the $\delta$-Sct type. This classification was also confirmed by \cite{Chen2022}.

\begin{figure}[h]
\centering
\includegraphics[width=\columnwidth]{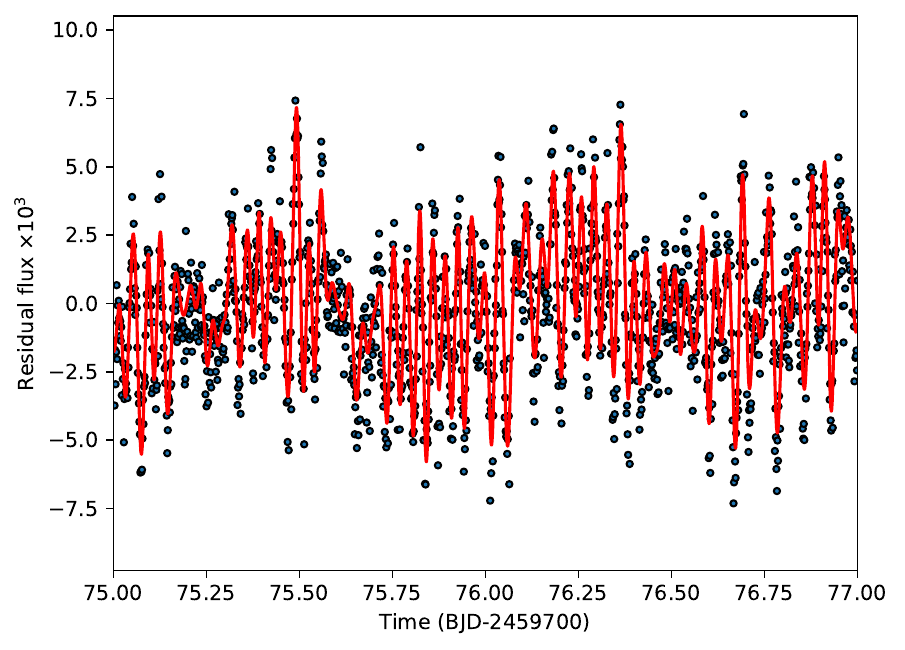}
\caption{Pulsations on residual LC from TESS after removing the model LC (Sec.~\ref{model}). The solid line shows the model from period analysis (70 strongest frequencies used).}
\label{fig:puls}
\end{figure}

We used 2-min cadence PDCSAP data obtained by the \textit{TESS} mission in sectors S14, S40, S41, and S54. The Nyquist frequency is 360 days$^{-1}$ (corresponding period of 4 min), which is sufficient to detect nearly all reasonable pulsations. The Nyquist frequency for long-cadence data from the \textit{Kepler} telescope is 24 days$^{-1}$ (period of one hour); therefore, it is not suitable for the analysis of short-periodic signals. A generalized Lomb-Scargle \citep[GLS;][]{Zechmeister2009} periodogram was used to find all frequencies in the data. A GLS periodogram was constructed from the residual data obtained after fitting the EB model (Sec.~\ref{model}). A part of the used LC is shown in Fig.~\ref{fig:puls}. Repeated pre-whitening was used to reduce the aliases of the already detected frequencies.

\begin{figure}[h]
\centering
\includegraphics[width=\columnwidth]{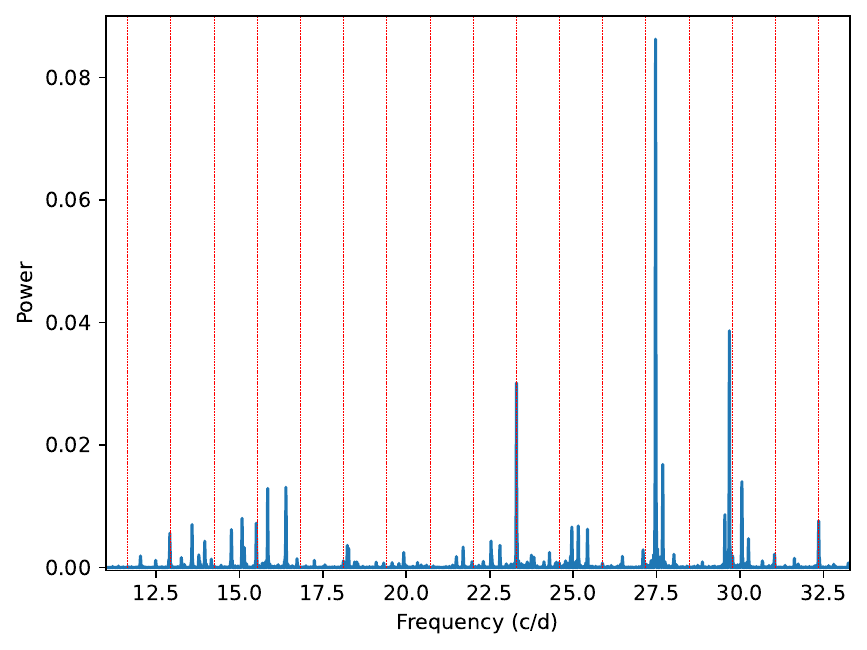}
\caption{Periodograms associated to the \textit{TESS} data. Red dashed lines represent multiplies of orbital frequency.}
\label{fig:gls}
\end{figure}

\begin{table*}[h]
\caption{Short-periodic signals (period less than 2 hours) found in the TESS data. Frequencies close to the multiple of the orbital one are excluded.}         
\label{tab:puls}    
\centering  
{\scriptsize \begin{tabular}{ccc}
	\hline\hline
	  Frequency    &      Period       & Amplitude  \\
	    (c/d)      &       (min)       &   (mmag)   \\ \hline
	27.471065 (07) & \,\,52.41879 (01) & 1.278 (14) \\
	29.683733 (11) & \,\,48.51143 (01) & 0.855 (13) \\
	30.058921 (15) & \,\,47.90591 (03) & 0.536 (12) \\
	27.687191 (21) & \,\,52.00960 (04) & 0.531 (12) \\
	16.390832 (15) & \,\,87.85398 (09) & 0.495 (12) \\
	15.845888 (15) & \,\,90.87530 (09) & 0.496 (12) \\
	15.077718 (17) & \,\,95.50516 (10) & 0.387 (11) \\
	13.577414 (19) &  106.05849 (14)   & 0.363 (11) \\
	25.152985 (19) & \,\,57.24966 (04) & 0.356 (11) \\
	24.959056 (22) & \,\,57.69449 (04) & 0.348 (11) \\
	29.548485 (18) & \,\,48.73346 (03) & 0.344 (11) \\
	14.757707 (19) & \,\,97.57613 (13) & 0.342 (11) \\
	25.430310 (24) & \,\,56.62534 (06) & 0.335 (10) \\
	30.253749 (25) & \,\,47.59740 (04) & 0.296 (10) \\
	22.539608 (25) & \,\,63.88753 (07) & 0.292 (10) \\
	13.960241 (23) &  103.15008 (17)   & 0.281 (10) \\
	18.232176 (24) & \,\,78.98125 (10) & 0.259 (10) \\
	22.807586 (26) & \,\,63.13689 (07) & 0.254 (10) \\
	21.703679 (25) & \,\,66.34820 (07) & 0.251 (10) \\
	18.271088 (25) & \,\,78.81304 (12) & 0.246 (10) \\
	15.144938 (28) & \,\,95.08127 (17) & 0.234 (10) \\
	27.095877 (29) & \,\,53.14461 (06) & 0.221 (09) \\
	19.921737 (26) & \,\,72.28286 (09) & 0.217 (09) \\
	28.018525 (37) & \,\,51.39456 (07) & 0.210 (09) \\
	24.291445 (34) & \,\,59.28013 (09) & 0.206 (09) \\
	26.476793 (31) & \,\,54.38724 (06) & 0.190 (09) \\
	12.031189 (28) &  119.68891 (27)   & 0.189 (09) \\
	24.920234 (29) & \,\,57.78436 (07) & 0.187 (09) \\
	21.499325 (30) & \,\,66.97885 (10) & 0.186 (09) \\
	23.824953 (29) & \,\,60.44083 (07) & 0.183 (09) \\
	27.751176 (33) & \,\,51.88969 (06) & 0.180 (09) \\
	13.778264 (35) &  104.51245 (26)   & 0.180 (09) \\
	23.748388 (33) & \,\,60.63569 (09) & 0.178 (09) \\
	13.258032 (33) &  108.61340 (27)   & 0.178 (09) \\
	31.633004 (35) & \,\,45.52207 (04) & 0.167 (09) \\
	16.727019 (33) & \,\,86.08825 (17) & 0.165 (09) \\
	14.154710 (34) &  101.73292 (24)   & 0.152 (09) \\
	17.244914 (37) & \,\,83.50288 (19) & 0.150 (09) \\
	25.087383 (37) & \,\,57.39936 (09) & 0.147 (09) \\
	22.310271 (42) & \,\,64.54427 (12) & 0.145 (09) \\
	12.489862 (36) &  115.29350 (33)   & 0.145 (09) \\
	27.593462 (39) & \,\,52.18628 (07) & 0.143 (09) \\
	27.682249 (35) & \,\,52.01889 (07) & 0.140 (09) \\
	13.805762 (47) &  104.30427 (36)   & 0.137 (09) \\
	30.677105 (40) & \,\,46.94054 (06) & 0.136 (09) \\
	23.622486 (34) & \,\,60.95887 (09) & 0.136 (09) \\
	19.097131 (39) & \,\,75.40400 (16) & 0.135 (09) \\
	15.102431 (44) & \,\,95.34888 (27) & 0.130 (09) \\
	18.520195 (39) & \,\,77.75296 (16) & 0.132 (09) \\
	12.869184 (43) &  111.89520 (37)   & 0.130 (09)
\end{tabular}
\begin{tabular}{ccc}
	\hline\hline
	  Frequency    &      Period       & Amplitude  \\
	    (c/d)      &       (min)       &   (mmag)   \\ \hline
	24.773933 (46) & \,\,58.12561 (12) & 0.128 (09) \\
	24.499933 (38) & \,\,58.77567 (09) & 0.129 (08) \\
	16.368815 (40) & \,\,87.97216 (22) & 0.128 (08) \\
	18.447763 (61) & \,\,78.05824 (26) & 0.126 (08) \\
	22.568005 (44) & \,\,63.80715 (13) & 0.126 (08) \\
	29.999250 (47) & \,\,48.00120 (07) & 0.124 (08) \\
	20.336735 (39) & \,\,70.80782 (13) & 0.124 (08) \\
	19.575305 (43) & \,\,73.56207 (16) & 0.121 (08) \\
	15.738589 (39) & \,\,91.49486 (23) & 0.119 (08) \\
	28.873326 (47) & \,\,49.87302 (09) & 0.118 (08) \\
	33.246809 (51) & \,\,43.31242 (07) & 0.118 (08) \\
	19.323322 (43) & \,\,74.52134 (16) & 0.117 (08) \\
	24.127530 (55) & \,\,59.68286 (13) & 0.116 (08) \\
	27.466122 (40) & \,\,52.42823 (07) & 0.114 (08) \\
	29.682744 (40) & \,\,48.51304 (07) & 0.112 (08) \\
	23.453270 (51) & \,\,61.39869 (13) & 0.111 (08) \\
	23.138561 (46) & \,\,62.23378 (13) & 0.110 (08) \\
	19.776155 (46) & \,\,72.81497 (17) & 0.110 (08) \\
	30.214568 (57) & \,\,47.65913 (09) & 0.109 (08) \\
	31.746594 (47) & \,\,45.35919 (07) & 0.108 (08) \\
	15.683232 (54) & \,\,91.81781 (32) & 0.108 (08) \\
	24.468390 (42) & \,\,58.85145 (10) & 0.108 (08) \\
	19.953908 (56) & \,\,72.16632 (20) & 0.106 (08) \\
	25.395442 (52) & \,\,56.70308 (12) & 0.105 (08) \\
	24.879345 (46) & \,\,57.87933 (10) & 0.105 (08) \\
	16.159788 (47) & \,\,89.11008 (26) & 0.102 (08) \\
	18.542661 (53) & \,\,77.65875 (22) & 0.100 (08) \\
	32.398479 (39) & \,\,44.44654 (06) & 0.100 (08) \\
	17.560432 (47) & \,\,82.00254 (22) & 0.099 (08) \\
	13.997625 (60) &  102.87459 (45)   & 0.098 (08) \\
	27.617636 (53) & \,\,52.14060 (10) & 0.097 (08) \\
	21.389330 (58) & \,\,67.32328 (19) & 0.097 (08) \\
	24.831177 (49) & \,\,57.99161 (12) & 0.096 (08) \\
	23.570904 (52) & \,\,61.09227 (13) & 0.095 (08) \\
	32.810422 (66) & \,\,43.88849 (09) & 0.091 (08) \\
	23.776156 (77) & \,\,60.56487 (20) & 0.091 (08) \\
	20.441788 (51) & \,\,70.44394 (17) & 0.090 (08) \\
	25.116051 (60) & \,\,57.33386 (13) & 0.090 (08) \\
	22.997652 (54) & \,\,62.61509 (14) & 0.088 (08) \\
	31.352894 (56) & \,\,45.92877 (09) & 0.087 (08) \\
	20.608848 (49) & \,\,69.87290 (16) & 0.087 (08) \\
	26.426379 (55) & \,\,54.49101 (12) & 0.087 (08) \\
	14.448390 (59) & \,\,99.66508 (40) & 0.086 (08) \\
	27.685484 (46) & \,\,52.01281 (09) & 0.084 (08) \\
	13.348078 (60) &  107.88071 (49)   & 0.083 (08) \\
	28.725587 (73) & \,\,50.12952 (13) & 0.083 (08) \\
	18.245027 (75) & \,\,78.92562 (33) & 0.082 (08) \\
	30.975099 (60) & \,\,46.48896 (09) & 0.082 (08) \\
	32.656213 (58) & \,\,44.09574 (07) & 0.082 (08) \\
	24.972356 (51) & \,\,57.66376 (12) & 0.081 (08)
\end{tabular}}
\tablecomments{Only 100 strongest signals are listed. Uncertainties are given in parentheses.}
\end{table*}

We detected a cluster of pulsation periods ranging from half an hour to two hours (see Fig.~\ref{fig:gls}). Table~\ref{tab:puls} lists 100 of the most significant ones. Their amplitudes go from 1.3~mmag to less than 0.1~mmag. Few of the detected signals had longer periods. However, they are all higher multiples of the orbital frequency or aliases corresponding to the technical parameters of \textit{TESS} observations (e.g. the duration of the sectors).

The shorter parts of the LC (five-day intervals similar to Sec.~\ref{spot} or the individual \textit{TESS} sectors) were analysed to study the possible time evolution of the pulsations. We did not observe any significant changes in the periods or amplitudes of detected signals over time.
%Some weak signals were not identified above the set false-alarm probability (FAP) level at these short intervals, which were insufficient for their detection. 

The range of detected pulsation frequencies and their large amounts confirm the previous classification of the pulsator as the $\delta$-Sct type \citep{Breger2000}. According to the estimated spectral types of the EB components, we could strongly assume that the primary star is the pulsating one. With an effective temperature of 7460~K and luminosity of about 13~L$_\sun$, it is located in the instability strip of the Hertzprung-Russell diagram, where the occurrence of pulsations is highly probable \citep{Murphy2019}. Moreover, we suggest that the star pulsates in the $p$~modes based on the detected frequencies of approximately 20~--~30~c/d, which are more typical for this mode of pulsations \citep{Kurtz2022}.

One could propose pulsation beats as an explanation for the changes in the O-C diagram and observed deformation of the LC (O'Connell effect) analysed in Sec.~\ref{lc}. Two close frequencies could significantly affect the shape of the LC depending on whether they would interfere in the same or opposite phases. However, we did not find any pair of frequencies that could generate a beat effect with a period close to the observed one on the O-C diagram (approximately 200~--~300 days).

For stars pulsating in radial modes, there is a simple relation between the pulsation period $P$ (in days) and the mean density of the star $\bar{\rho}$ usually written in the form of
\begin{equation}
    P=Q\left( \frac{\bar{\rho}}{\rho_\sun} \right)^{-0.5},
\end{equation}
where $Q$ is the pulsation constant that has different values for different pulsation modes. Assuming $M\approx1.8$~M$_\sun$ and $R\approx 2.18$~R$_{\sun}$ (Tab.~\ref{tab:all-params}) and dominant frequency 27.47~c/d ($P=0.0364$~days), we obtain $Q\approx0.015$ which corresponds to the 3$^{\rm rd}$ radial overtone \citep{Poro2024}.

%We used the \textsc{Python} package \textsc{Lightkurve} to perform simple asteroseismology analysis \citep{Lightkurve}. The obtained frequency of the maximum oscillation amplitude $\nu_{\rm max}$ is 321.5~$\mu$Hz (27.78~c/d), and the average frequency spacing $\Delta\nu$ is 25.58~$\mu$Hz (2.21~c/d).
%A regular frequency spacing recently detected in tens of $\delta$~Sct pulsators \citep{Bedding2020} allows us to estimate the physical parameters of stars using scaling relations \citep[e.g.][]{Chaplin2013,Aerts2021}, assuming that the oscillations are similar to those in our Sun. However, the observed pulsations of KIC\,7023917 are not of the solar-type and the possible regularity of the frequency spectrum is very weak and of a random nature.

\section{Discussions and Conclusions}
\label{conc}

In our study, we present an analysis of a close detached eclipsing binary KIC\,7023917. We used data from \textit{Kepler} and \textit{TESS} space missions. In addition, we performed ground-based photometry using Sloan filters $g'$ and $r'$ and radial velocity measurements.

The times of eclipses show quasi-periodic changes that are antisymmetric for primary and secondary ones. Their amplitudes varied from 5 to 10 min and the period was approximately 200--300 days. Similar variations were observed in the shapes of the maxima with changes in the maxima amplitude of up to 5~mmag. We found a significant correlation between the changes in the O-C diagram and the O'Connell effect. The presence of a cold star spot on the secondary component can describe these features. Its size varies from 0 to 25\degr, whereas a larger spot causes a lower height of primary maxima (around phase 0.25), a later determined time of primary minima, and the opposite behaviour for secondary ones.

The resolution and quality of the obtained spectra were insufficient to detect the spectral lines of the secondary component. The semi-amplitude of the measured RV is 22~km/s, which provides a mass function of $10^{-3}$~M$_\sun$. 
%The spectroscopic orbit is nearly circular which confirms also LC analysis.

Ground-based observations were used to estimate the colour indices and effective temperature of the primary component. The resulting value is 7587~K, and it is slightly higher than that given in the \textit{Gaia} DR3 database. 
%The major source of this difference is the quality of ground-based data strongly affected by the local weather conditions.
%The model parameters of EB are comparable with the values determined from \textit{Kepler} and \textit{TESS} data but with significantly bigger uncertainties caused by the worse quality of the input data.

LC modelling determined the effective temperature of the secondary component to be approximately 6400K (spectral type F6~--~F7) compared to the temperature of the primary component of 7460~K (type A7). The orbital inclination of the system is 58.5\degr. The mass ratio is very low, only 0.1, which is surprising for the pair of stars of the determined spectral types. However, this value is in very good agreement with the RV measurements. This unusually low mass ratio can be explained by the past mass transfer from the secondary component to the primary one. This hypothesis is also supported by the fact that the secondary star nearly fills its Roche lobe in the current system geometry.

Using the semi-amplitude of RV or the distance from \textit{Gaia} as two partially independent methods, we were able to determine the absolute parameters of the system and its components. The semi-major axis is approximately 4.5~R$_\sun$. The radii of the stars are 2.2~R$_\sun$ and 0.9~R$_\sun$. Their masses are 1.8~M$_\sun$ and 0.2~M$_\sun$. The corresponding luminosities are 13.2~L$_\sun$ and 1.2~L$_\sun$.

\begin{figure}[h]
\centering
\includegraphics[width=\columnwidth]{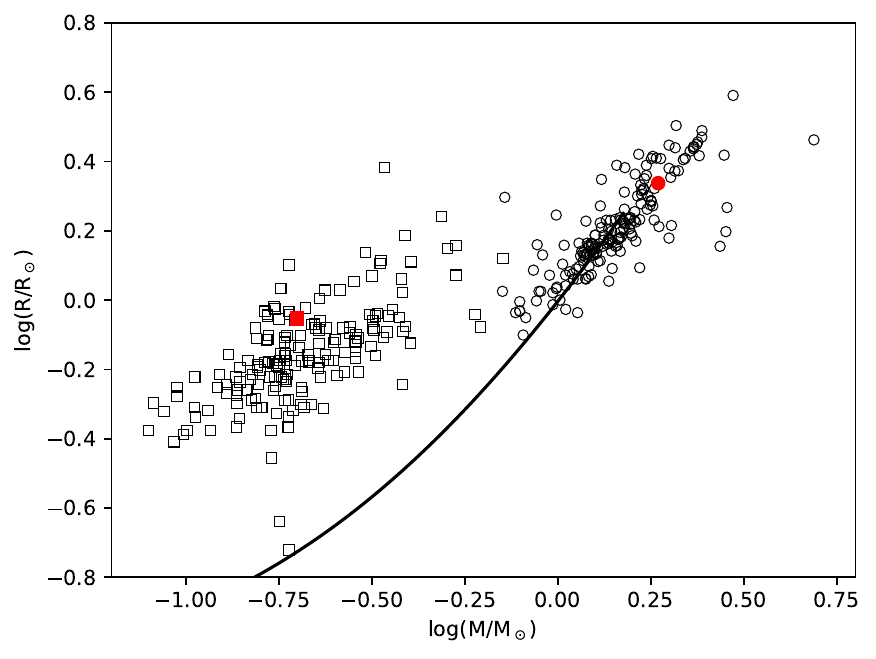}
\caption{The mass--radius diagram for primary (red circle) and secondary component (red square) of KIC\,7023917. White symbols show similar low-mass EBs from literature \citep[both over-contact and detached systems;][and references]{Christopoulou2022}. The solid line represents the empirical relation for low-mass main sequence stars \citep{Eker2018}.}
\label{fig:mr}
\end{figure}

With the determined set of stellar parameters, KIC\,7023917 belongs to a unique group of eclipsing binaries with unusually low mass ratio \citep{Yang2015,Christopoulou2022}. The primary components are main sequence stars and low-mass secondary ones are evolved stars with super-luminosities and large radii as a result of extensive mass transfer in binary evolution history. Moreover, components of similar systems could merge and be a progenitor of a single rapid-rotating star, maybe blue stragglers, as already discussed by some authors \citep[e.g.][]{Eggleton2001,Liu2024,Yang2015}. However, most of these low-mass ratio binaries are overcontact. Only a few detached binaries with low mass ratios are known -- e.g. KIC\,8262223 \citep{Guo2017} or KIC\,10661783 \citep{Southworth2011}. The mass--radius diagram (Fig.~\ref{fig:mr}) confirms that the parameters of both components of the studied system are typical for similar low-mass ratio binaries with evolved secondary components significantly above the main sequence. The secondary star is slightly more oversized than the typical one. 

From the evolutionary point of view, the initial mass of the secondary component was bigger than the primary one. Using empirical relation for binaries after extended mass-transfer phase \citep{Yildiz2013,Liu2024}, we estimated its initial mass to 2~--~2.5~M$_\sun$ and corresponding spectral type late B or early A. The initial mass of the primary one was probably lower than 1~M$_\sun$. Therefore, the mass ratio was reversed. However, these values should be considered as only very approximate and illustrative ones. The more massive component evolved faster. As it expanded, it filled its Roche lobe and the mass transfer started. Now, the secondary is a star which finished mass transfer and starts to contract from its current oversized dimensions \citep{Guo2017}. Its effective temperature will rise with decreasing radius while keeping nearly constant luminosity. The terminal state will be a pre-helium white dwarf in the typical EL~CVn-type binary \citep[e.g.][]{Peng2024}. The radius of a final white dwarf (with an effective temperature of about 10,000 -- 20,000~K) will be probably less than 0.1~--~0.2~R$_\sun$.

The data from \textit{TESS} show an additional short-periodic signal. On the periodogram, we detect the strongest peak with frequency 27.5~c/d (period 52.4 minutes and amplitude 1.3~mmag) and a large amount of other weaker but still significant periods in the range of 30~--~120 minutes. These periods are probably the result of $\delta$-Scuti pulsation of the primary component of the EB in the $p$~mode. The dominant detected frequency corresponds to the 3$^{\rm rd}$ radial overtone. The parameters of the primary component are typical for the $\delta$-Scuti pulsator. However, these pulsations could be partially started or driven by the evolutionary history of binary and past mass transfer \citep{Wang2020}. 

\begin{acknowledgements}
This article is based on the data collected with the Perek's 2-m telescope. We are grateful to the operators of Perek's telescope (Radek Novotn\'{y}, Lud\v{e}k \v{R}ezba, Ji\v{r}\'{i} Srba and Eva \v{Z}\v{d}\'{a}rsk\'{a}) for help in obtaining the spectroscopic observations. 

This work was supported by the Slovak Research and Development Agency under contract No. APVV-20-0148 and by The Ministry of Education, Youth and Sports of the Czech Republic by grant LTT-20015. The research of P.G. was supported by the internal grant No. VVGS-2023-2784 of the P. J. {\v S}af{\'a}rik University in Ko{\v s}ice. 
\end{acknowledgements}

\facilities{Gaia, Kepler, OO:2, SuperWASP, TESS}

\software{ELISa \citep{Cokina2021}, iSpec \citep{Blanco2014,Blanco2019}, lightkurve \citep{Lightkurve}, OCFit \citep{Gajdos2019}, PyAstronomy \citep{Czesla2019}}

\bibliographystyle{aasjournal}
\bibliography{bibfile}

\end{document}